\documentclass[aps,prl,reprint,groupedaddress]{revtex4-1}
\usepackage{graphics,graphicx}

\begin{document}


\title{Strong Lensing with Finite Temperature Scalar Field Dark Matter}

 \author{V. H. Robles\footnote{Part of the Instituto Avanzado de Cosmolog\'ia (IAC)
  collaboration http://www.iac.edu.mx/}}
 \email{vrobles@fis.cinvestav.mx}
 \author{T. Matos}
 \email{tmatos@fis.cinvestav.mx}
 \affiliation{Departamento de F\'isica, Centro de Investigaci\'on y de Estudios Avanzados del IPN, A.P. 14-740, 07000 M\'exico D.F.,
             M\'exico.}



\begin{abstract}
We investigate the gravitational constraints imposed to dark matter halos in the context of finite temperature 
scalar field dark matter. We find constraints to produce multiple images by dark matter only, we show that there 
are differences with respect to the full Bose Einstein condensate halo when the temperature of the scalar field in dark matter 
halos is taken into account. The non zero temperature allows the scalar field to be in excited states and recently, their 
existence has proved to be necessary to fit rotation curves of dark matter dominated galaxies of all sizes, it 
also explained the non universality of the halo density profiles. Therefore, we expect that combining observations of rotation 
curves and strong lensing systems can give us a clue to the nature of dark matter. Finally, we propose a method to identify 
the excited state of a strong lens halo, knowing various halo excited states can provide information of the scalar field dark matter 
halo evolution which can be tested using numerical simulations.

\end{abstract}


\maketitle


\textit{Introduction.-} In the $\Lambda$ Cold Dark Matter ($\Lambda$CDM) model, known as the standard model of cosmology, 
the formation of structures in the universe is through a hierarchical process of growth of structures, meaning that small structures,
like small halos of galaxies merge to form bigger ones, like galaxy clusters and superclusters halos. 
The $\Lambda$CDM model can successfully describe cosmological observations 
such as the large scale distribution of galaxies, the temperature variations in the 
cosmic microwave background radiation and the recent acceleration of the universe 
\citep{a1,a2}.
However, recent observations in far and nearby galaxies have shown that the model 
faces some conflicts at small scales, the galaxy rotation curves are more 
consistent with a constant central density \citep{a1,a3,a4} than a cusp, known as the cusp/core problem \citep{a4}.

One model that has been widely discuss is the Scalar Field Dark Matter model (SFDM). A review of the model can be seen in 
\cite{a5}. Recently, it has been shown to agree with rotation curves (RC) of dwarf, small 
low surface brightness (LSB) galaxies \cite{a6,a7}, and large LSB galaxies \cite{a8}. 
The main idea is that the dark matter (DM) is a fundamental spin-0 scalar field $\Phi$ with a repulsive interaction, 
in \cite{a8} the model was extented to include the dark matter temperature and finite temperature corrrections 
up to one-loop in the perturbations.

Besides the RCs, the gravitational lensing offers a way to differentiate halo profiles. While the reconstruction 
of the mass profile in cluster size halos is usually more challenging than in galaxy size halos, we can still obtain valuable
constraints from the strong lensing by galaxy size halos. For strong lensing systems, usually elliptical galaxies, 
in \cite{a9,a10} they showed that the the total mass profiles have steeper inner slopes 
than a Burkert profile \cite{a11}, however, it remained inconclusive if it was due to a distribition of baryons or 
a cuspy halo. In \cite{a12} the halos of elliptical galaxies were modeled using a NFW profile and showed that 
there was an excess of concentration to explain the observed gravitational lensing, and in \cite{a13} they give 
a condition to produce strong lensing when the dark matter behaves as a Bose Einstein condensate (BEC), in this case, the 
dark matter is a spin-0 scalar field at temperature zero and in which the bosons are all in the ground state. 
They applied their condition to a subsample of strong lensed systems in SLACS\cite{a14} and CASTLES\cite{a15} and showed 
that for these systems the product $\rho_c r_{max}$, with $\rho_c$ the central density and $r_{max}$ the halo radius, is at 
least one order of magnitude greater than the one found in dwarf 
galaxies and small LSB galaxies \cite{a6,a16}, as in a BEC halo the $r_{max}$ is constant, they conclude that the 
those strong lensing systems must be denser than dwarf galaxies. 
From all these results, it is clear that the combination of strong lensing and RCs in galaxies are able to constraint the 
different models of DM and can be the key to find the real nature of the DM in the universe.
 
In this work we provide the strong lensing constraints for the SFDM model when the dark matter temperature is taken into account.
In this model DM halos can be in excited states because now the temperature T$\neq$0, this solves discrepancies in 
rotation curve fits found when T=0, for instance, the presence of a constant halo radius ($r_{max}$) for all galaxies and in the case of 
large galaxies, the incapability to fit at the same time the inner and outermost regions of RCs, see \cite{a8} for details. 
Our results modify the conclusion in \cite{a13} as now, the halo radius can be different. 

Additionally, using the exact solution of a static SFDM configuration found in \cite{a8} to model a DM halo, 
we discuss how multiple images produced by a galaxy can be accounted for with dark matter only, mainly because different 
DM central mass concentrations, high and low, can be accommodated in the SFDM model in the form of different excited states of the SF. 
We also describe a mechanism to identify the state of the SFDM halos that satisfy the strong lensing constraints.
 
 \ \

{\it Finite temperature SFDM lens equation}.- 
The formation of DM halos in the SFDM model with non zero temperature was described in \cite{a8}. Essentially, 
the dark matter is an auto-interacting real scalar field (SF) in a thermal bath at temperature T with an initial $Z_2$ symmetric potential, 
as the universe expands and cools, the temperature drops until it reaches a critical temperature so that the $Z_2$ symmetry is 
spontaneously broken and the field rolls down to a new minimum where it oscillates. The critical temperature T$_C$,
determines the moment in which the DM fluctuations can start growing, they do it from the moment when T$<$T$_C$ until they 
reach a stable equilibrium point. When the SF is near the minimum of the potential and after the SB, 
a spherically symmetric static SFDM halo has a temperature-corrected density profile given by 
\begin{equation}
\rho (r) = \rho_{0} \frac{\sin^{2} (kr)}{(kr)^{2}},  \label{density}   
\end{equation}
where $\rho_{0}$ is the central density, $k=\pi j/R$, $R$ is defined by the condition $\rho(R)=0$,
and $j$ is the number of the minimum excited state required to fit a galaxy RC up to the last measured point.
We can compare it with the zero-temperature profile of a BEC halo \cite{a7} 
\begin{equation}
\rho^0 (r) = \rho_{c} \frac{\sin (\pi r/r_{max})}{(\pi r/r_{max})}.  \label{temp0density}   
\end{equation}
To obtain the lens equation for the SFDM model we follow the procedure of \cite{a13}. We consider the thin lens 
aproximation and the weak field limit, under this hyphotheses the lens equation for a spherically symmetric mass 
distribution can be written 
\begin{equation}
 \beta=\theta - \frac{M_p(\theta)}{\pi D_{OL}^2 \theta \Sigma_{crit}} \label{beta}
\end{equation}
where $\beta$ and $\theta$ are the positions of the source and the image, respectively. $\Sigma_{crit}\equiv c^2 D_{OS}/4 \pi G D_{OL}D_{LS}$,
$D_{OL}$, $D_{OS}$, $D_{LS}$ are the distances between observer and the lens, observer and the source, and from the lens to the source.
The projected mass $M_p(\xi)$ enclosed within a projected radius $\xi$ is:
\begin{equation}
M_p(\xi_{\ast})=  \frac{4 \rho_0 R^3}{\pi j^2} \int_0^{\xi_{\ast}} \xi'_{\ast} d\xi'_{\ast} \int_0^{\sqrt{1-\xi'^2_{\ast}}}
\frac{\sin^2(\pi \sqrt{\xi'^2_{\ast}+z^2_{\ast}})}{\xi'^2_{\ast}+z^2_{\ast}}dz_{\ast}  \label{masa}
\end{equation}
where $0 \leq \xi_{\ast}\equiv \xi/R \leq 1$ and $z_{\ast} \equiv z/R$. Defining $\beta_{\ast} \equiv D_{OL}\beta/R$, 
$m(\xi_{\ast})\equiv M_p(\xi_{\ast})/\rho_0 R^3$, and $\theta_{\ast} \equiv \theta D_{OL}/R$, the dimensionless lens equation:
\begin{equation}
 \beta_{\ast}(\theta_{\ast})= \theta_{\ast} -\lambda_T \frac{m(\theta_{\ast})}{\theta_{\ast}} \label{dimbeta}
\end{equation}
with
\begin{equation}
 \lambda_T= \frac{\rho_0 R}{\pi \Sigma_{crit}} = 0.57 h^{-1} \frac{\rho_0}{M_{\odot} \ pc^{-3}} \frac{R}{kpc}\frac{1}{f_{dist}} 
\label{lambda}
\end{equation}
with $f_{dist}\equiv d_{OS}/d_{OL}d_{LS}$ a distance factor, and the reduced angular distance $d_A \equiv D_A H_0 /c$. $H_0 \equiv 100h km s ^{-1}Mpc^{-1}$ 
is the Hubble constant today and $h=0.71$ \cite{a17}.

 \begin{table}
 \caption{\label{table1}Minimum values of $\lambda_T$ to produce strong lensing for $j$=1,2,3,4,5. $j$=0 corresponds to a zero temperature halo.}
 \begin{ruledtabular}
 \begin{tabular}{lcccccc}
 j & 0 & 1 & 2 & 3 & 4 & 5 \\
 $\lambda_{T,min}$  & 0.27 & 0.35 & 0.66 & 0.98 & 1.32 & 1.63 
 \end{tabular}
 \end{ruledtabular}
 \end{table}

 \begin{figure}
\centering
 \resizebox{70mm}{!}{\includegraphics{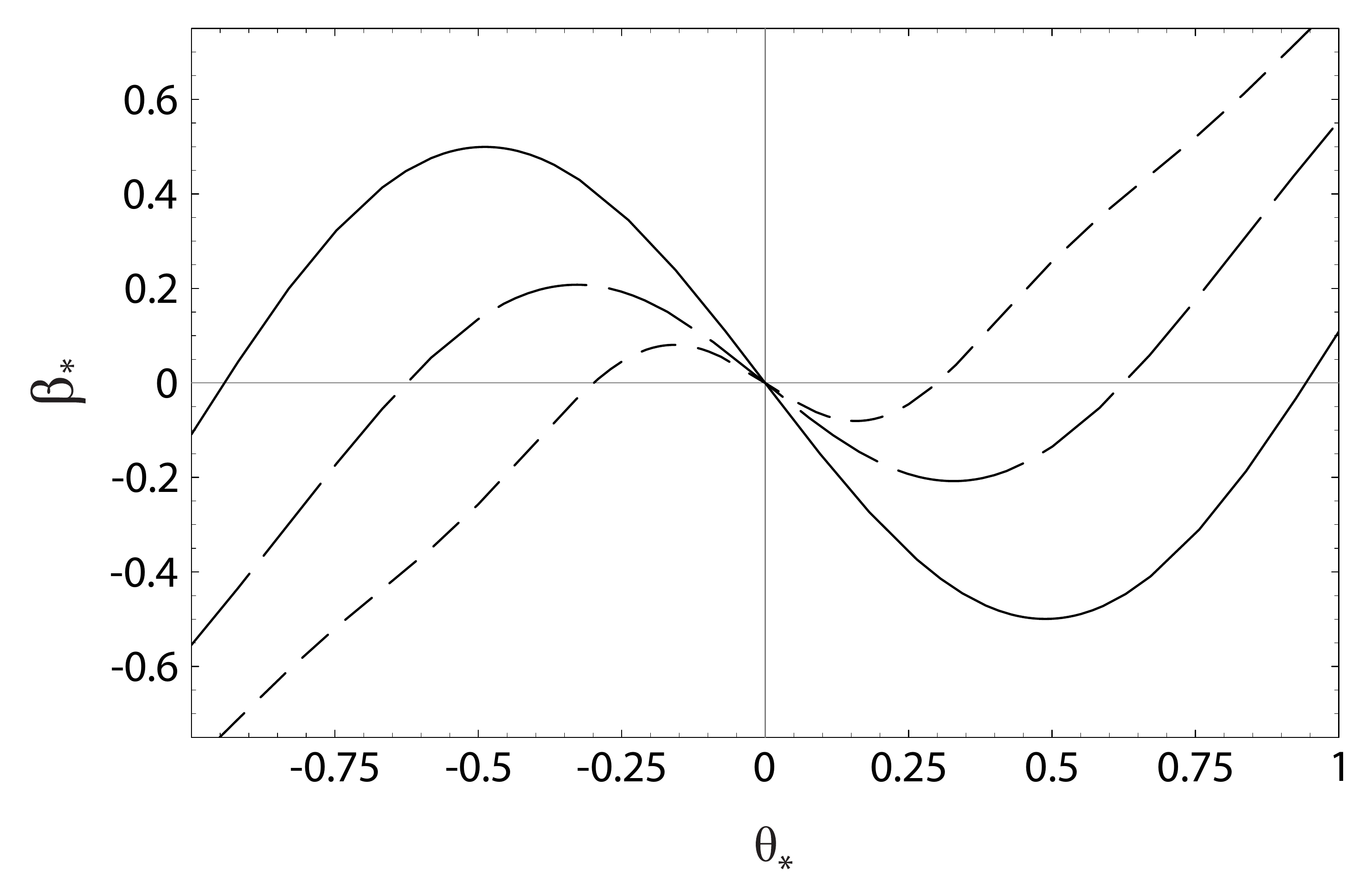}}%
 \caption{\label{figura1}Lens equation of the finite temperature SFDM halo for the states j=0,1,2. The intersection with the horizontal 
  axis define $\theta_{\ast,E}$. The BEC halo at zero temperature is denoted as j=0. $\lambda_T=0.7$ for j=0,1, and $\lambda_T=1.2$ for 
 j=2. The values are chosen above their minimum value to produce strong lensing. The solid line represents the BEC halo, the 
 j=1,2 finite temperature SFDM halos correspond to the long-dashed and small-dashed lines, respectively.}
 \end{figure}

 \begin{figure}
\centering
  \resizebox{75mm}{!}{\includegraphics{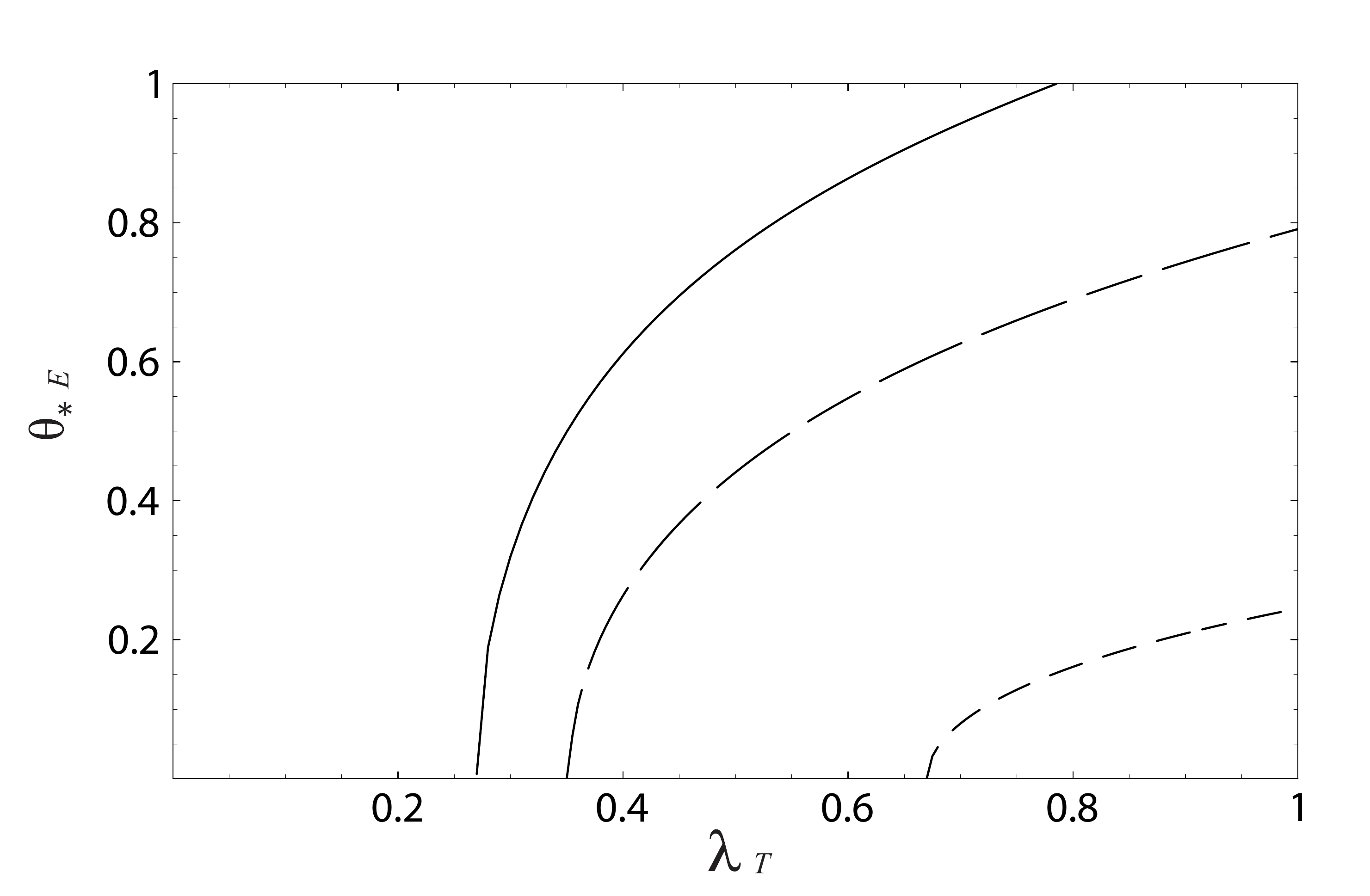}}%
 \caption{\label{figure2} Einstein radius $\theta_{\ast,E}$ as function of $\lambda_T$. Shown are the BEC halo (solid line), 
  j=1 (long-dashed line), and j=2 (small dashed line) finite temperature SFDM halos.}
 \end{figure}

 \ \

{\it Results and Discussion}.- 
The parameter $\lambda_T$ determines the properties of the SFDM profile lens. In Fig. 1 we show the lens equation for $j=1,2$, and 
the BEC halo (we define this case as $j$=0). The values of $\lambda_T$ in Fig. 1 are all chosen to be above their minimum value to produce 
multiple images, only those configurations with impact parameter $|\beta_{\ast}| < \beta_{\ast c}$ can produce strong lensing. 
$\beta_{\ast c}$ is the maximum of $\beta_{\ast}(\theta_{\ast})$ in Fig.1 for $\theta_{\ast}<0$. In Table 1 we report the minimum 
values of $\lambda_T$ for the first five exited states. 
From Table 1 we notice that the minimum value of $\lambda_T$ depends on the exited state of the SF in the halo, larger values 
are required for higher states, also, when the temperature is taken into account $\lambda_{T,min}>\lambda_{0,min}$ for all $j$, 
where $\lambda_{0,min}$ is the minimum value to produce strong lensing for the BEC halo at $T$=$0$. Comparing with the value given
in \cite{a13}, we see that for $j$=$1$ $\lambda_{T,min}$=$0.35>0.27$=$\lambda_{0,min}$, although it is a small difference, the 
real advantage of adding temperature to the SFDM is the capability to fit a broader variety of density profiles and not only 
dwarfs. 
From Eq. (\ref{lambda}) we can rewrite the condition to produce strong lensing with a finite temperature SFDM halo as 
\begin{equation}
 \rho_0 R (M_{\odot} pc^{-2}) \geq (1754.38) \lambda_{T,min} h f_{dist} \label{slenscondition}
\end{equation}

In Fig. 2 we plot the Einstein radius for the BEC halo and halos with $j$=1,2. This caustic point corresponds 
to $\beta(\theta_{\ast,E})=0$.
An important effect of the inclusion of temperature is that $\theta_{\ast,E}$ is smaller for higher values of $j$ for 
the same value of $\lambda_{T}$.
This allows us to determine the excited state of the halo form observations. For instance, if an observation of a strong 
lens system gives us a value of $\lambda_{T}$ higher than the minimum for $j$=2, from Fig.2 we see that the strong lensing 
could be due to a SFDM halo in the first or second state, however, their Einstein radius are different, therefore, observing 
the Einstein radius in these systems can tell us the state in which the halo is found. Applying this method 
to a wide sample of strong lensing systems would allow us to extract information on the scalar field halos, like 
their stability or the existence of correlations between $j$ and redshift, which can be compared with cosmological simulations.

In the case of a BEC halo, the radius $r_{max}$ is a constant, therefore, if we replace $\rho_0$ by $\rho_c$ and $R$ by 
$r_{max}$ in Eq. (\ref{slenscondition}) we obtain the constraint found in \cite{a13}, where they conclude that 
their sample of strong lensing systems are at least ten times denser than dwarf systems. In our case $R$ is a parameter and is 
not a constant for all halos \cite{a8}. Thus, we have two posible limits. One limit happens when the lens systems have similar 
radius than dwarfs or small LSBs, the conclusion for this case is the same as in \cite{a13}, the lensed halos are at least 10 
times denser than these small systems, something that should be expected as these low density systems do not produce strong lensing. 
The other possibility is that the lens halos are 10 times larger and have similar densities than small systems, and an intermediate situation could be 
one bewteen the two previous cases. Thus, only the observations of the halos will decide the case to which they correspond. 

The condition to produce multiple images with only dark matter is given form Table 1, if it is found that for a strong lensing 
system $\lambda_{T}< 0.35$ then the various images can not be due to a SFDM halo, in this case it could be the baryons that cause 
the lensing. 
We also notice from Eq. (\ref{slenscondition}) and Table 1, that if $\lambda_{T}> 0.35$ then strong lensing can be achieved at least 
with $j$=$1$, for higher values of $\lambda_{T}$ more excited states are allowed to produce multiple images. However,
we have discussed how we can differentiate the state of the SF in a halo using $\theta_{\ast,E}$, this angle is related to the halo mass 
distribution which is a function of $j$. 

It can be seen from Eq. (\ref{density}) that for a fixed halo radius, as $j$ increases the density 
distribution shifts towards the center of the halo, the higher the $j$, the more the central region contributes to the total 
mass and the less the contribution from the outer regions. Thus, for high $j$, $\theta_{\ast,E}$ is closer to the center of the 
SFDM halo. If we combined this case with the constraint equation (eq. (\ref{slenscondition})), then, increasing the excited states 
implies that the dark matter is more concentrated in the center in order to have strong lensing. Therefore, a high central mass concentration can be 
obtained with a SFDM halo in a exited state and multiple images can form with only DM.
This can also explain dark matter dominated galaxy clusters that present multiple images 
and that have a core behavior in their centers as discussed in \cite{a18,a19}. 
Additionally, multiple image systems in which an apparent cuspy DM halo is necessary can be explained with the SFDM model 
using an adequate exited state of a SFDM halo, in these cases strong lensing can also be atributed to DM. 
The posibility that a high central density is due to baryons is not discarded, however, as there are a variety of 
systems with different central densities, luminosities etc., the explanation of this non-univesality of 
profiles can be well described within the SFDM model and it is not essential to know all the baryonic physics behind.

 \ \
 
{\it Conclusions}.- In this work we have given a strong lensing constraint of a finite temperature scalar field DM halo. We extended the previous
analysis of a fully condensed system at temperature zero, and show that multiples images are posible with 
more than one state of the scalar field. As finite temperature DM halos are not of only one radius, then, our constraint 
expresses two limits, either the halos of strong lensing systems are 10 times larger or 10 times denser than dwarf 
galaxy halos. We also provide a way to identify the excited state of the DM halo in these systems by means of measuring their 
Einstein radius, the closer it is to the center the more probable that the SFDM halo is in a higher excited state. 
A deeper analysis of this can be used as a test to the validity of the SFDM model, mainly, because 
identifying the excited states of various halos could give us information about their evolution which 
can be compared with simulations and work as test to our model. 

We would like to point out that based on galactic observations, there seems to be a trend that low central density 
galaxies possess SFDM halos in the base state, while higher density galaxies prefer higher states, for the latter the baryons 
could accrete faster towards the center producing higher central baryonic concentrations and changing the shape of the inner profile, 
thus, there could be a tight relation between the type of galaxy and the state of the scalar field in their DM halos, however 
numerical simulations will be needed to verify these remarks.

Finally, it is important to notice that our finite temperature density profile can describe strong lensing without the need of high 
quantities of baryons in the center, a high density concentration in the central region of a galaxy halo can be explained 
with excited states of the SF, and for the clusters of galaxies we can have core profiles to the very center due to the form of 
our DM density profile that is always core, although in these complex systems a superposition of states could be necessary, mainly 
because of their complex evolution that could have changed substantially the DM halo profile.
In the case that the baryonic concentration is non negligible the central DM halo distribution could be affected, however, a detail 
analysis of the interaction bewteen DM and baryons will be given in a future work. 

 \ \
{\it Acknowledgments}.- This work was partially supported by CONACyT M\'{e}xico
under grants CB-2009-01, no. 132400, CB-2011, no. 166212,  and I0101/131/07 C-
234/07 of the Instituto Avanzado de Cosmologia (IAC) collaboration
(http://www.iac.edu.mx/). V.H.R. is supported by a CONACYT scholarship.



\begin{thebibliography}{91}
\bibitem[\protect\citeauthoryear{Coles}{}]{a1} P. Coles, Nature \textbf{433}, 248 (2005).
\bibitem[\protect\citeauthoryear{Guo et al.}{}]{a2} Q. Guo et al., MNRAS \textbf{413}, 101 (2011).
\bibitem[\protect\citeauthoryear{Kuzio de Naray et al.}{2008}]{a3} R. Kuzio de Naray, S.S. McGaugh, and W. de Blok, ApJ \textbf{676}, 920 (2008).
\bibitem[\protect\citeauthoryear{de Blok}{2010}]{a4} W. de Blok, Adv. Astron. 2010, Article ID 789293, 14 pages (2010).
\bibitem[\protect\citeauthoryear{Suarez et al.}{2013}]{a5} A. Su\'{a}rez, V.H. Robles, and T. Matos, arXiv:1302.0903 (2013).
\bibitem[\protect\citeauthoryear{Robles and Matos}{2012}]{a6} V. H. Robles and T. Matos, Mon. Not. R. Astron. Soc. \textbf{422}, 282 (2012).
\bibitem[\protect\citeauthoryear{Bohmer \& Harko }{2007}]{a7} C.G. B\"{o}hmer and T. Harko, JCAP06 (2007), 025 2007.
\bibitem[\protect\citeauthoryear{Robles and Matos}{2013}]{a8} V. H. Robles and T. Matos, ApJ \textbf{763}, 19 (2013).
\bibitem[\protect\citeauthoryear{Yousin}{2003}]{a9} Y. Park and H.C. Ferguson, ApJ \textbf{589}, L65 (2003).
\bibitem[\protect\citeauthoryear{Koopmans}{2003}]{a10} L.V.E Koopmans and T. Treu, ApJ \textbf{583}, 606 (2003).
\bibitem[\protect\citeauthoryear{Burkert}{1995}]{a11} A. Burkert, Astrophys. J. \textbf{447}, L25 (1995).
\bibitem[\protect\citeauthoryear{Keeton}{2001}]{a12} C.R. Keeton, ApJ \textbf{561}, 46 (2001).
\bibitem[\protect\citeauthoryear{Alma}{2013}]{a13} A.X. Gonzalez-Morales, A. Diez-Tejedor, L.A. Ure\~{n}a-Lopez,and O. Valenzuela, 
Phys. Rev. D \textbf{87}, 021301(R) (2013). 
\bibitem[\protect\citeauthoryear{Bolton}{2008}]{a14} A.S. Bolton, S. Burles, L.V. Koopmans, T. Treu, R. Gavazzi, L.A. Moustakas, R. Wayth, and D.J. Schlegel,
Astrophys. J. \textbf{682}, 964 (2008).
\bibitem[\protect\citeauthoryear{Falco}{}]{a15} E. Falco, C. Kochanek, J. Lehar, B. McLeod, J. Munoz et al., arXiv:astro-ph/9910025, http://www.cfa.harvard
.edu/castles/.
\bibitem[\protect\citeauthoryear{Harko}{2011}]{a16} T. Harko, J. Cosmol. Astropart. Phys. 05 (2011) 022
\bibitem[\protect\citeauthoryear{Jarosik}{2011}]{a17} N. Jarosik, C. Bennett, J. Dunkley, B. Gold, M. Greason et al., Astrophys. J. Suppl. Ser. \textbf{192}, 14 (2011).
\bibitem[\protect\citeauthoryear{Newmann}{2011}]{a18} A.B. Newman, T. Treu, R.S. Ellis, and D.J. Sand, ApJ \textbf{728} L39 (2011).
\bibitem[\protect\citeauthoryear{Laporte}{2012}]{a19} C.F.P. Laporte, S.D.M. White, T. Naab, M. Ruszkowski, and V. Springel, MNRAS, \textbf{424}, 747 (2012).

\end{thebibliography}
\end{document}